\documentclass[aps,preprint,amsmath,amssymb,citeautoscript,showpacs]{revtex4}

\usepackage{graphicx,times}
\usepackage{longtable}

\begin{document}

\title{Finite size effects in the presence of a chemical potential: \\
A study in the classical non-linear $O(2)$ sigma-model} 
\author{Debasish Banerjee$^1$, Shailesh Chandrasekharan$^{1,2}$}
\email{debasish@theory.tifr.res.in,sch@phy.duke.edu}
\affiliation{
$^1$Tata Institute of Fundamental Research, Homi Bhabha Road, Mumbai 400005 India\\
$^2$Department of Physics, Box 90305, Duke University, Durham, North Carolina 27708, USA.}

\begin{abstract}
In the presence of a chemical potential, the physics of level crossings leads to singularities at zero temperature, even when the spatial volume is finite. These singularities are smoothed out at a finite temperature but leave behind non-trivial finite size effects which must be understood in order to extract thermodynamic quantities using Monte Carlo methods, particularly close to critical points. We illustrate some of these issues using the classical non-linear $O(2)$ sigma model with a coupling $\beta$ and chemical potential $\mu$ on a $2+1$ dimensional Euclidean lattice. In the conventional formulation this model suffers from a sign problem at non-zero chemical potential and hence cannot be studied with the Wolff cluster algorithm. However, when formulated in terms of world-line of particles, the sign problem is absent and the model can be studied efficiently with the ``worm algorithm''. Using this method we study the finite size effects that arise due to the chemical potential and develop an effective quantum mechanical approach to capture the effects. As a side result we obtain energy levels of up to four particles as a function of the box size and uncover a part of the phase diagram in the $(\beta,\mu)$ plane. 
\end{abstract}

\pacs{11.10.-z, 02.70.Ss, 05.30.-d, 75.10.Hk}

\maketitle

\section{Introduction}

Understanding the phase diagram of quantum chromodynamics (QCD) as a function of temperature $T$ and baryon chemical potential $\mu$ is an active area of research. Although much is known about the physics at $\mu=0$ from lattice QCD calculations \cite{Bazavov:2009bb,DeTar:2009ef}, there is controversy of what might occur at non-zero $\mu$ and small values of $T$ \cite{Hatsuda:2009uu,RevModPhys.80.1455,McLerran:2007qj}. Due to the sign problem, which arises in all current formulations of lattice QCD at non-zero $\mu$, it is impossible to perform first principles calculations to settle the controversy today. Most of our knowledge of the $(T,\mu)$ phase diagram of QCD is based on models that are motivated from universality and solved using mean field theory. Can at least some of these models be studied from first principles? For example, recently a Landau-Ginzburg approach was used to uncover parts of the phase diagram of QCD where the low energy physics is described by bosonic excitations \cite{Hatsuda:2006ps}. In these regions it should be possible to construct bosonic effective field theory models that share the same symmetries, low energy physics and possibly the phase transitions as QCD. It would be interesting to study these models from first principles. Unfortunately, sign problems also arise in bosonic field theories in the presence of a chemical potential when formulated in the conventional approach. For this reason not many first principles studies of field theories with a chemical potential exist. However, many of these sign problems are solvable today and thus allow us to explore the physics of a chemical potential from first principles. It may be useful to study these simpler field theories before attempting to study QCD.

One of the simplest examples of a relativistic bosonic field theory is the classical non-linear $O(2)$ sigma model on a cubic lattice which has been studied extensively in the context of superfluid transitions using the efficient Wolff cluster algorithm \cite{Wolff:1988kw}. The phase transition is between two phases: an $O(2)$ symmetric phase and a phase where the symmetry is spontaneously broken. Close to the phase transition the low energy physics is described by an interacting quantum field theory of massive charged bosons in the symmetric phase and of massless Goldstone bosons in the broken phase. At the critical point the low energy physics is scale invariant and the critical behavior belongs to the three dimensional $XY$ universality class. 

Since the model contains an exact $O(2)$ global symmetry, one can also introduce a chemical potential $\mu$ that couples to the corresponding conserved charge. This chemical potential helps one study the ``condensed matter'' composed of the fundamental boson present in the theory. When $\mu \neq 0$, the action in the conventional formulation becomes complex and Monte Carlo algorithms suffer from a sign problem exactly like in QCD. Not surprisingly, the phase diagram of the condensed matter arising in the $O(2)$ non-linear sigma model has not been studied from first principles. On the other hand non-relativistic bosonic lattice models, especially in the Hamiltonian formulation have been studied for many years by the condensed matter community. Here one naturally constructs the field theory with bosonic world lines and there is no sign problem when one introduces a chemical potential. Thus, it is natural that a world-line approach could solve the corresponding sign problem even for a relativistic field theory. This was shown explicitly for both the linear sigma model \cite{Endres:2006xu} and the $O(2)$ non-linear sigma model \cite{Chandrasekharan:2008gp}. 

While the world-line representation for bosonic lattice field theories was well known for many years, the main advance in the field that improved our ability to perform a first principles calculation in the presence of a chemical potential, was the discovery of an efficient Monte Carlo algorithm called the ``worm algorithm'' \cite{Prokof'ev:2001zz}. Variants of this algorithm in the name of ``directed loop algorithm'' \cite{Syljuasen:2002zz,Adams:2003cc} have been used to solve a variety of models that arise in the strong coupling limit of lattice gauge theories \cite{Karsch:1988zx,Chandrasekharan:2006tz,Chandrasekharan:2007up,deForcrand:2009dh}. The worm algorithm has also been found to be an efficient approach to study a wider class of fermionic field theories in the loop representation in two dimensions where fermion sign problems are absent \cite{Wolff:2008km,Wenger:2008tq,Wolff:2008xa,Wolff:2009kp} and weak coupling Abelian lattice gauge theory \cite{Azcoiti:2009md}. A combination of the worm algorithm and the determinantal algorithm was recently developed to solve the lattice Thirring model in the fermion bag formulation in higher dimensions \cite{Chandrasekharan:2009wc}. All these developments should allow us to explore the physics of a chemical potential using first principles in a variety of lattice models with interesting symmetries.

In this work, we explore the $O(2)$ non-linear sigma model in the presence of a chemical potential and show that interesting finite size effects naturally arise due to the level crossing phenomena. Understanding these effects is important to extract the thermodynamic limit and thus uncover the $(\beta,\mu)$ phase diagram, where $\beta$ is the coupling and $\mu$ is the chemical potential. Our studies should be useful for future studies since the finite size effects we uncover is a universal feature. Our work also provides accurate results that can be used to compare with results from other methods, like the complex Langevin method, which are being explored as a solution to the sign problem in general \cite{Aarts:2008wh,Aarts:2009dg,Guralnik:2009pk}. Our work is organized as follows: In section \ref{model} we discuss our model and observables in order to set up the notation. In section \ref{fse} we discuss the finite size effects that arise in the presence of a chemical potential and develop an effective quantum mechanical description that captures these effects in section \ref{effqm}. Sections \ref{res1} and \ref{res2} contain our results obtained using the worm algorithm. In particular we show that the observed finite size effects are described well by the effective quantum mechanical description of section \ref{effqm}. In section \ref{pdiag} we discuss the phase diagram of the $O(2)$ model which emerges from our work.

\section{Model and Observables}

\label{model}

The action of the $O(2)$ non-linear sigma model on a lattice with a finite chemical potential that we study here is given by
\begin{equation}
S = -\beta \sum_{x,\alpha} 
\Big\{ \cos(\theta_x - \theta_{x+\alpha} - i\mu \delta_{\alpha,t}) \Big\},
\end{equation}
where $x$ is the lattice site on a three dimensional cubic lattice, $\alpha =1,2$ represent the spatial directions and $\alpha=t$ represents the temporal direction. We will use $L$ to represent the spatial size and $L_t$ the temporal size and assume periodic boundary conditions. The constant $\beta$ plays the role of the coupling. The chemical potential $\mu$ is introduced in the standard way and couples to the conserved charge of the global $O(2)$ symmetry \cite{Hasenfratz:1983ba}. When $\mu \neq 0$ the action becomes complex and Monte Carlo algorithms to generate configurations $[\theta]$ that contribute to the partition function
\begin{equation}
Z = \int [d\theta_x] \mathrm{e}^{-S},
\end{equation}
suffer from a sign problem. In particular the Wolff cluster algorithm \cite{Wolff:1988kw} is no longer useful at non-zero chemical potential. Hence the phase diagram of the model in the $(\beta,\mu)$ plane remains unexplored.

It is possible to avoid the sign problem if we rewrite the partition function in the world-line representation \cite{Chandrasekharan:2008gp}. Using the identity
\begin{equation}
\exp \left\{ \cos \theta \right\} = \sum_{k=-\infty}^{\infty}I_k(\beta) e^{ik\theta},
\end{equation}
where $I_k$ is the modified Bessel function of the first kind, on each bond $(x,\alpha)$, and performing the angular integration over $\theta_x$ the partition function can be rewritten as
\begin{equation}
\label{wlpart}
Z = \sum_{[k]} \ \ \prod_{x} \ \ 
\Big\{I_{k_{x,\alpha}}(\beta) e^{\mu \delta_{\alpha,t} k_{x,\alpha}}\Big\}\ \ 
\delta \Big(\sum_{\alpha} (k_{x,\alpha}-k_{x-\alpha,\alpha})\ \Big),
\end{equation}
where the bond variables $k_{x,\alpha}$ describe ``world-lines'' or  ``current'' of particles moving from lattice site $x$ to the site $x+\hat{\alpha}$ and take integer values. A configuration of these bond variables, denoted by $[k]$, is thus a world-line configuration.  The global $U(1)$ symmetry of the model is manifest in the local current conservation relation represented by the delta function. In other words any particle that comes into the site must leave the site due to current conservation. In this world-line formulation the partition function is a sum over explicitly positive terms even in the presence of $\mu$.  Details of the ``worm algorithm'' that we have developed to update the world-line configuration $[k]$ is described in appendix \ref{algo}.

We focus on four observables in this work:
\begin{enumerate}
\item The average particle density $\rho$:
\begin{equation}
\rho = \frac{1}{L^2} \langle \sum_{x\in \mathrm{timeslice}} k_{x,t} \rangle
\end{equation}
The average particle number is then given by $\langle N \rangle = \rho L^2$.
\item The particle number susceptibility $\kappa$:
\begin{equation}
\kappa = \frac{1}{L^2L_t} \langle \left( \sum_{x} k_{x,t}\right)^2 \rangle
\end{equation}
Note that $\kappa = L_t/L^2 \langle N^2 \rangle$.
 \item The superfluid density (or particle current susceptibility) $\rho_s$:
\begin{equation}
\rho_s = \frac{1}{2L^2L_t} \langle \sum_{\alpha=1,2} \left( \sum_{x} k_{x,\alpha}\right)^2 \rangle
\end{equation}
The superfluid density is known to be $\rho_s = 1/L_t \langle W^2\rangle$ where $W$ is the spatial winding number of particles \cite{PhysRevB.39.2084}. We define $\langle N_s \rangle = L^2 \rho_s$ as the number of particles that are in the superfluid phase in a finite system.
\item The condensate susceptibility $\chi$:
\begin{equation}
\chi = \sum_{y} \langle \mathrm{e}^{i\theta_x}\mathrm{e}^{-i\theta_y} \rangle
\end{equation}
\end{enumerate}
The first three observables are ``diagonal'' observables and can be measured on each world-line configuration and then averaged. The condensate susceptibility $\chi$ on the other hand is a ``non-diagonal'' observable, but it can be related to the size of each worm update as discussed in appendix \ref{algo}. We discuss some tests of the algorithm in appendix \ref{algotest}. In particular we have been able reproduce earlier results of the $O(2)$ non-linear sigma model at $\mu=0$. One of these results is the estimate of the critical coupling $\beta_c = 0.45421$ \cite{Hasenbusch:1989fi}. For $\beta > \beta_c$ the $O(2)$ symmetry is spontaneously broken, while for $\beta < \beta_c$ the model is in the symmetric phase. Our tests also show agreement with exact calculations on a $2 \times 2$ lattice at non-zero $\mu$.

\section{Finite Size Effects}
\label{fse}

\begin{figure}[t]
\begin{center}
\vskip0.5in
\includegraphics[width=0.7\textwidth]{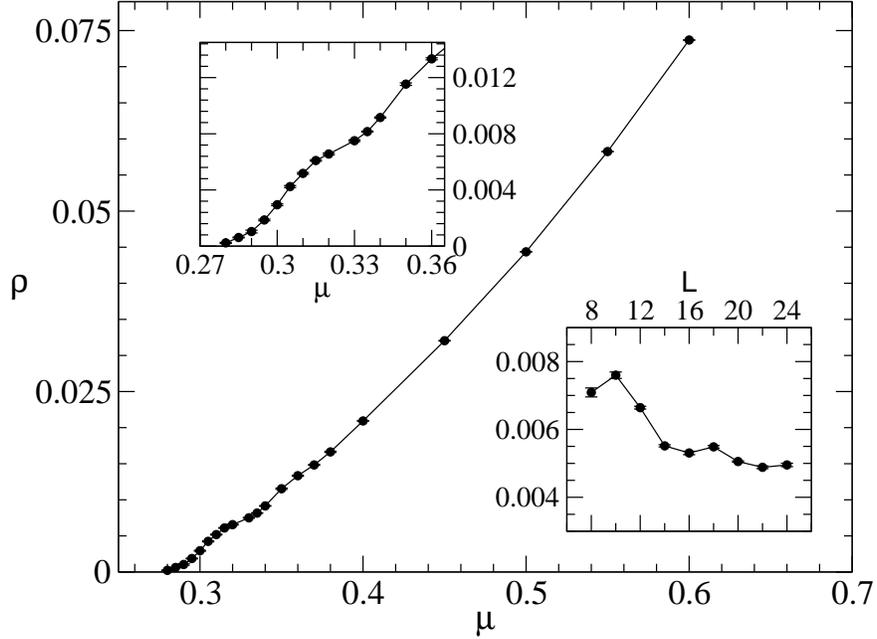}
\end{center}
\caption{\label{fig1} Plot of $\rho$ as a function of $\mu$ for $\beta=0.43$, $L=12$ and $L_t=144$. The data shows clear ``wiggles'' at small values of $\rho$ which disappears for larger values. The left inset magnifies the region of the ``wiggles''. The right inset shows the plot of $\rho$ as function of $L$ at $\mu=0.32$ assuming $L_t=L^2$ which also shows clear non-monotonic behavior.}
\end{figure}

A good understanding of finite size effects is important for extracting thermodynamic results from numerical computations. This is particularly true close to a second order critical point where correlation lengths diverge. While developing a theory of finite size effects, one usually assumes $L_t = L^z$ where $z$ is called the dynamical critical exponent of the problem. Such a choice makes calculations natural. In a relativistic theory since $z=1$, it is natural to choose $L_t=L$. Most studies of the critical behavior at $\beta=\beta_c$ and $\mu = 0$ make use of this choice. On the other hand, in the presence of a chemical potential, since the low effective energy theory is non-relativistic, one expects $z=2$ and $L_t = L^2$ is a more natural choice. However, in our work we have found that even with this choice the finite size effects close to the critical point are non-trivial in the presence of a chemical potential. In fact observables always show clear ``wiggles'' and cannot be fit to a simple power law that one expects near the critical point. In order to demonstrate this feature, in Fig.~\ref{fig1} we plot the behavior of the average particle density $\rho$ as a function of $\mu$ for $L=12$ and $L_t=144$ at $\beta=0.43$. From mean field theory we expect 
\begin{equation}
\rho \approx \left\{\begin{array}{cc} 
c(\mu - \mu_c) & \mu > \mu_c \cr 
0 & \mu < \mu_c
\end{array}\right.
\end{equation}
close to $\mu=\mu_c$ in the thermodynamic limit. In Fig.~\ref{fig1} we observe that $\rho$ is indeed zero for $\mu < 0.27$ and begins to increase for $\mu \gtrsim 0.27$. But the increase, although roughly linear close to $\mu_c$ as expected, shows clear ``wiggles'' when $0.27 < \mu < 0.38$ and only for $\mu > 0.38$ the ``wiggles'' disappear. The region between $0.27 < \mu < 0.38$ has been enlarged in the left inset in order to enhance the observed ``wiggles''. In the right inset we fix $\mu=0.32$ and plot $\rho$ as a function of $L$ assuming $L_t=L^2$. Again the data shows clear oscillations whose origin may seem a bit mysterious. In order to avoid these oscillations one has to go to much larger $L$ at a fixed value of $\mu$. However, since $L_t$ scales like $L^2$ going to larger lattice sizes is more difficult than in a relativistic theory. For this reason, we believe it may be useful to develop a different type of finite size analysis. 

As we will argue below, the strange finite size behavior is the result of energy levels crossing each other due to the chemical potential. Since the particle number is a conserved quantum number, energy levels with different particle numbers can cross each other at critical values of the chemical potential. Similarly at a fixed chemical potential, the changes in spatial size can also cause these energy levels to cross each other. These level crossings lead to singularities at low temperatures (large $L_t$) in a finite spatial volume (fixed $L$). While these singularities are smoothed out at finite $L_t$, they leave behind non-trivial finite size effects observed above. While it may still be possible to develop practically useful finite size scaling relations using $L_t=L^2$, we find it natural to consider a finite size scaling theory for quantities as a function of $L_t$ and $\mu$ for a fixed value of $L$ close to the critical values of $\mu$ where energy levels cross each other. As we discuss below this leads to an effective quantum mechanics problem. The finite size effects studied here have been observed earlier in the context of quantum spin-systems in a magnetic field \cite{PhysRevE.66.046701}, but they were not analyzed using the techniques we introduce below.

\section{Effective Quantum Mechanics}
\label{effqm}

At a fixed value of $L$ for sufficiently large $L_t$, it must be possible to map the lattice field theory problem to an effective quantum mechanics problem where only a few low energy levels play an important role. Let us label these energy levels by $|N,k\rangle$ and the energy eigenvalues by $E_k^{(N)}$, where $N=0,1,2,...$ represents the particle number sector of the energy level and $k$ represents ``other'' quantum numbers. The levels and the energies depend on $L$ and $\beta$. The partition function of the problem may be written as
\begin{equation}
\label{Pf}
Z = \sum_{k,N} \mathrm{e}^{-(E_k^N - \mu N)L_t}
\end{equation}
Using this effective quantum mechanical description we can in principle find the $L_t$ and $\mu$ dependence of various quantities. However, for the analysis to be practically useful we need to assume that only a few energy levels are important. If we assume that $\mu$ is close to a critical value $\mu_c$ where level crossing phenomena occurs, then for large enough $L_t$ one might expect the physics to be dominated by just two levels. In this approximation we will derive the $L_t$ and $\mu$ dependence of all our observables.

In a given particle number sector can assume $E_0^{(N)} < E_1^{(N)} < E_2^{(N)} < ...$ without loss of generality. However, in this work we will also assume that $E_0^{(0)} < E_0^{(1)} < E_0^{(2)}...$ which means that it always costs energy to add a particle into the system. While this is not necessary it is  precisely the situation we encounter in this work and simplifies our analysis. With these assumptions it is easy to argue that close to the critical chemical potential where the particle number changes from $N$ to $N+1$ we can approximate the partition function to be
\begin{equation}
Z \approx \mathrm{e}^{-(E_0^{(N)}-\mu N)L_t} + \mathrm{e}^{-(E_0^{(N+1)}-\mu (N+1))L_t}
\end{equation}
Here we have assumed all higher energy states will be suppressed exponentially at large $L_t$. It is easy to verify that $\mu_c^{(N)} \equiv E_0^{(N+1)}-E_0^{(N)}$ is the critical chemical potential where the average particle number changes from $N$ to $N+1$. Below we discuss the $L_t$ and $\mu$ dependence of each observable when $\mu \approx \mu_c^{(N)}$.

\subsection{Particle Number}
We first consider the average particle number $\langle N\rangle$. When $\Delta_\mu^{(N)} = \mu - \mu_c^{(N)}$ is small and $L_t$ is large we can write
\begin{equation}
\label{avgN1}
\langle N \rangle = \frac{N + (N+1)\mathrm{e}^{\Delta_\mu^{(N)} L_t}}
{1 + \mathrm{e}^{\Delta_\mu^{(N)} L_t}}
\end{equation}
We will demonstrate later that our data fits very well to this simple one parameter fit and we are able to extract $\mu_c^{(N)}$ very accurately for all $L \leq 20$ for a variety of values of $\beta$.

\subsection{Number Susceptibility}
Next we discuss the number susceptibility $\kappa = L_t/L^2 \langle N^2 \rangle$. We now obtain
\begin{equation}
\langle N^2 \rangle = 
\frac{N^2 + (N+1)^2 e^{\Delta_\mu^{(N)} L_t}}{1 + e^{\Delta_\mu^{(N)} L_t}}
\end{equation}
The value of $\mu_c^{(N)}$ is the same as obtained from the average particle number. So this observable has no new free parameters.

\subsection{Superfluid Density}
In the effective quantum mechanical description the superfluid density is given by
\begin{equation}
\rho_s \ =\ \frac{1}{Z}\ \int_0^{L_t}\ dt \ 
\mathrm{Tr}\Bigg(\mathrm{e}^{-(L_t-t)\ H} O^\dagger \mathrm{e}^{-t\ H} O \Bigg)
\end{equation}
where
\begin{equation}
O = \frac{1}{L} \sum_{x_2} J_1(x_1,x_2)
\end{equation}
is an operator in the Hilbert space made up of the conserved current operator $J_i(x_1,x_2)$ in the direction $i$ at the site with coordinates $(x_1,x_2)$. Note the sum is over the surface perpendicular to the direction of the current. Since it is a conserved current it does not matter which surface one chooses. Now if we introduce a complete set of energy eigenstates we get
\begin{equation}
\label{2pt}
\rho_s = \frac{1}{Z}\ 
\sum_{n,k}\mathrm{e}^{-(E_k^{(n)}-n\mu)L_t}
\sum_{n'k'} |\langle n,k| O | n'k'\rangle|^2
\frac{\Big(1-\mathrm{e}^{-(E_{k'}^{(n')}-E_k^{(n)}-(n'-n)\mu)L_t}\Big)}
{(E_{k'}^{(n')}-E_k^{(n)}-(n'-n)\mu)}
\end{equation}
First we note that $\langle n,k | O | n'k'\rangle \propto \delta_{n n'}$, since the current operator commutes with the particle number operator and hence does not change the particle number. Further, as before we assume only two low lying energy levels are important in the partition function when $\mu \approx \mu_c^{(N)}$. Then the $[k,n]$ sum is replaced by $k=0$ and $n = N, N+1$. Hence we obtain
\begin{equation}
\rho_s = \frac{\rho_0 + \rho_1 \mathrm{e}^{\Delta_\mu^{(N)} L_t}}
{(1+\mathrm{e}^{\Delta_\mu^{(N)} L_t})}
\end{equation}
where 
\begin{equation}
\rho_0 = \sum_{k' \neq 0}  |\langle N,0| O | N k'\rangle|^2 
\frac{\Big(1-\mathrm{e}^{-\Delta E_{k'}^{(N)} L_t}\Big)}{\Delta E_{k'}^{(N)}},\ \ \ 
\mbox{with}\ \Delta E_{k'}^{(N)} \equiv E_{k'}^{(N)} - E_0^{(N)}.
\end{equation}
and $\rho_1$ is obtained by replacing $N$ with $N+1$ in the above expression. Note that the sum over $k'$ does not contain the $k'=0$ sum because $\langle N,0| O | N 0\rangle = 0$ since $O$ is a current operator and the ground state is rotationally invariant. Thus, the expression for $\rho_s$ contains two new parameters since $\mu_c^{(N)}$ has already been encountered before.

\subsection{Condensate Susceptibility}
The expression for the condensate susceptibility can also be obtained using eq.(\ref{2pt}) if the operator $O$ is replaced by
\begin{equation}
O = \frac{2}{L} \sum_x cos(\theta_{x,t}).
\end{equation}
The matrix element $\langle n,k | O | n'k'\rangle$ is non-zero only when $n'=n+1$ or $n'=n-1$. When $\mu \approx \mu_c^{(N)}$, the $[k,n]$ sum is again dominated by $E_0^{(N)}$ and $E_0^{(N+1)}$. However, in the present case the $\mu$ dependence also enters the $k'$ sum. In the limit as $\Delta_\mu^{(N)} \rightarrow 0$ and $L_t \rightarrow \infty$ the $k'=0$ term is singular while the other terms are not. Separating the singular term from others we find
\begin{eqnarray}
\chi &=& 2 |\langle N,0| O | N+1,0\rangle|^2
\frac{(\mathrm{e}^{\Delta_\mu^{(N)} L_t}-1)}
{\Delta_\mu^{(N)} (1+\mathrm{e}^{\Delta_\mu^{(N)} L_t})}
\nonumber \\
&& + \frac{1}{(1+\mathrm{e}^{\Delta_\mu^{(N)} L_t})}\ 
\sum_{k'\neq 0} 
|\langle N,0| O | N+1,k'\rangle|^2
\frac{(1-\mathrm{e}^{-(\Delta E^{(N+1)}_{k'}-\Delta_\mu^{(N)})L_t})}
{(\Delta E^{(N+1)}_{k'}-\Delta_\mu^{(N)})}
\nonumber \\
&& + \frac{\mathrm{e}^{\Delta_\mu^{(N)} L_t}}{(1+\mathrm{e}^{\Delta_\mu^{(N)} L_t})}\ 
\sum_{k'\neq 0} 
|\langle N+1,0| O | N,k'\rangle|^2
\frac{(1-\mathrm{e}^{-(\Delta E^{(N)}_{k'}+\Delta_\mu^{(N)})L_t})}
{(\Delta E^{(N)}_{k'}+\Delta_\mu^{(N)})}.
\end{eqnarray}
Since $|\Delta_\mu^{(N)}|$ is assumed to be much smaller than all $\Delta E^{(N)}_{k'}$ and $\Delta E^{(N+1)}_{k'}$, at large $L_t$ the exponentials in the $k'$ sum can be dropped. If the remaining terms are expanded in powers of $\Delta_\mu^{(N)}$ we find
\begin{equation}
\chi = \frac{\chi_0 (\mathrm{e}^{\Delta_\mu^{(N)} L_t}-1)/\Delta_\mu^{(N)} + 
(\chi_1  + \chi_2 \Delta_\mu^{(N)} + ...) + (\chi_1'+\chi_2'\Delta_\mu^{(N)} + ...)
\mathrm{e}^{\Delta_\mu^{(N)} L_t}}{(1+\mathrm{e}^{\Delta_\mu^{(N)} L_t})}.
\label{chiN}
\end{equation}
We find that our data fits well to this expression truncated at the quadratic order in $\Delta_\mu^{(N)}$, which means we have seven new parameters in our fit. However, most of these parameters are not determined reliably and contain large systematic errors. The only parameter that can be determined reliably is $\chi_0$ and this is what we quote as a result from our analysis.

\section{Results}
\label{res1}

We have performed extensive calculations at $\beta = 0.43, 0.50$ and $0.20$. These values of $\beta$ are chosen so that two of them are close to the critical coupling $\beta_c=0.45421$ on either side and one is far from it in the massive (disordered) phase. In this section we present fits of our results to the effective quantum mechanics description discussed above. As mentioned earlier, the effective description becomes useful only in the limit of small temperatures where excitations to higher energy levels can be neglected. Since the spacing between energy levels decreases with increase in volume, our approach works best on small spatial volumes. However, thanks to the efficient worm algorithm, we have been able to extract parameters of the effective quantum mechanics up to $L = 16$ at $\beta=0.43$ and $\beta=0.5$. Although this lattice size is small compared to normal studies of bosonic lattice field theories, it still allows us to perform a useful study of the $L$ dependence of the physics and draw quantitative conclusions about the thermodynamic limit. At $\beta=0.2$ we observe that the energy levels are more densely packed and we are able to compute quantities only up to $L=8$.

\begin{figure*}[t]
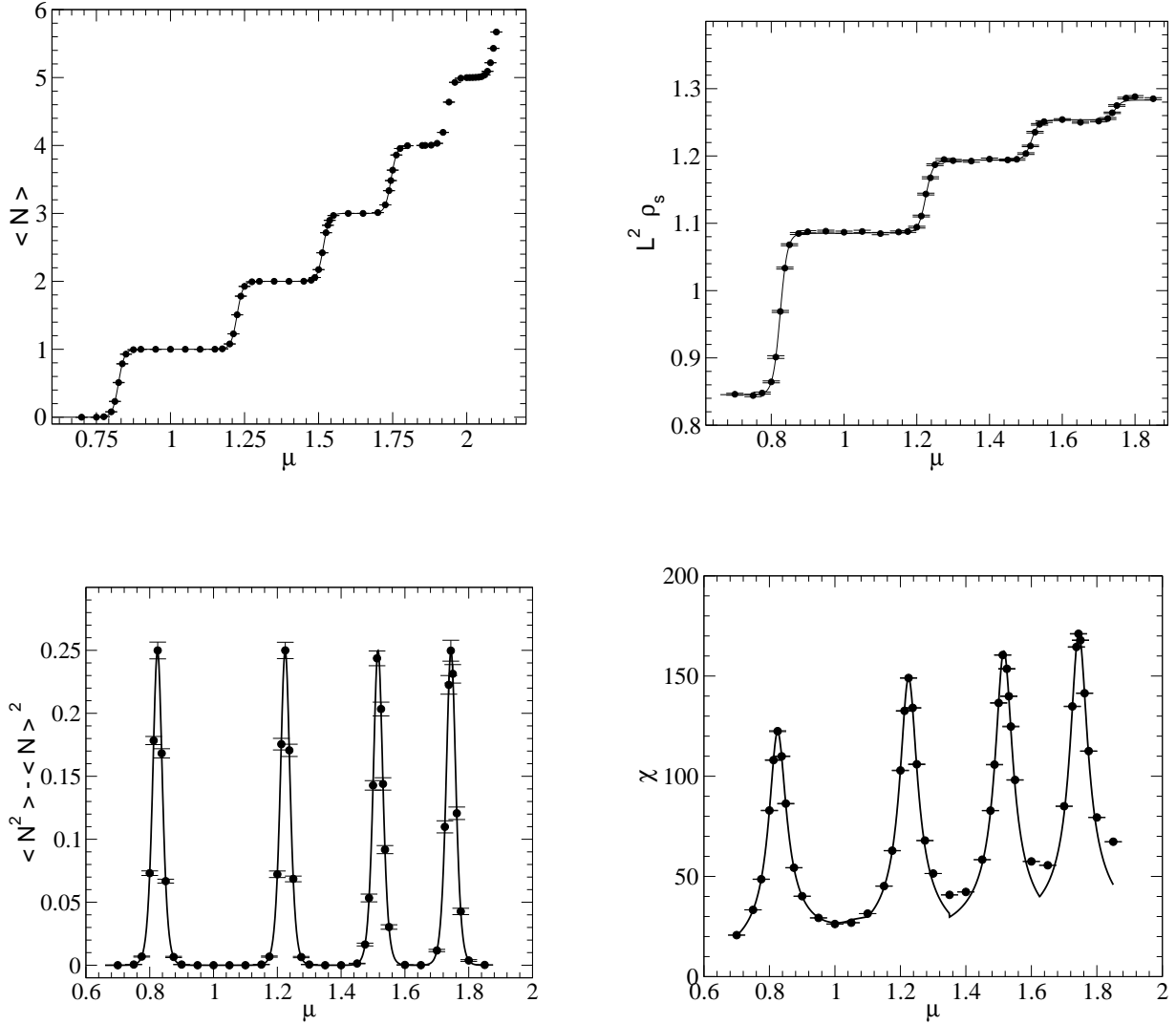

\begin{center}
\vbox{
\hbox{
\includegraphics[width=0.44\textwidth]{fig2a.eps}
\hskip0.5in
\includegraphics[width=0.46\textwidth]{fig2b.eps}
}
\vskip0.5in
\hbox{
\includegraphics[width=0.45\textwidth]{fig2c.eps}
\hskip0.5in
\includegraphics[width=0.45\textwidth]{fig2d.eps}
}
}
\end{center}
\caption{\label{fig2} The four observables as a function of $\mu$ up to four particle excitation. The data shown is for $L=2$, $L_t=100$ and $\beta=0.43$. The solid lines are fits to the effective quantum mechanical description.}
\end{figure*}

\begin{table}[t]
\begin{tabular}{|c|c|c|c|c|c|c|c|c|}
\hline
$N$ & $\mu^{(N)}_c$ & $\rho_0$ & $\rho_1$ & $\chi_0$ & \multicolumn{4}{|c|}{$\chi^2/DOF$}\\
\hline
\multicolumn{9}{|c|}{$\beta=0.43$} \\
\hline
0 & 0.82444(3) &  0.2114(2) & 0.2713(2) &  2.349(4) & 1.23 &  1.43 &  0.70 &  0.58 \\
1 & 1.22462(2) &  0.2715(2) & 0.2983(2) &  2.817(4) & 0.52 &  0.50 &  1.04 &  0.81 \\
2 & 1.51567(2) &  0.298(1) &  0.312(1)  &  3.013(4) & 0.73 &  0.70 &  1.20 &  1.28 \\
3 & 1.74436(2) &  0.313(1) & 0.321(1)   &  3.121(4) & 1.02 &  1.02 &  0.78 &  1.14 \\
\hline
\multicolumn{9}{|c|}{$\beta=0.50$} \\
\hline
0 & 0.63275(3) & 0.288(1) & 0.336(1) & 2.63(1)    & 0.58 & 0.56 & 1.33 & 0.68\\
1 & 1.05865(2) & 0.339(1)  & 0.362(1)  & 2.957(9) &  0.72 & 0.69 & 0.71 & 1.47\\
2 & 1.35787(2) & 0.364(1)  & 0.376(1)  & 3.116(8) & 0.89 & 0.85 & 1.37 & 0.94\\
3 & 1.58951(2) & 0.377(1)  & 0.386(1)  & 3.212(8) & 0.57 & 0.54 & 1.07 & 1.00\\
\hline
\multicolumn{9}{|c|}{$\beta=0.20$} \\
\hline
0 & 1.9141(1) & 0.0414(2) & 0.0884(2) & 1.458(9) & 0.55 & 0.52 & 0.47 & 0.81 \\
1 & 2.12263(9) & 0.0884(2) & 0.1106(3) & 2.14(1) &  0.41 & 0.40 & 1.38 & 0.98\\
2 & 2.33075(8) & 0.1109(2) & 0.1178(3) & 2.40(1) &  0.43 & 0.34 & 1.02 & 1.67\\
3 & 2.52196(8) & 0.1186(3) & 0.1150(2) & 2.48(1) &  0.56 & 0.52 & 1.36 & 1.17\\
\hline
\end{tabular}
\caption{\label{tab1} Parameters of effective quantum mechanics that describes the data for the $L=2$ lattice at $\beta=0.43$.}
\end{table}

\begin{figure*}[t]
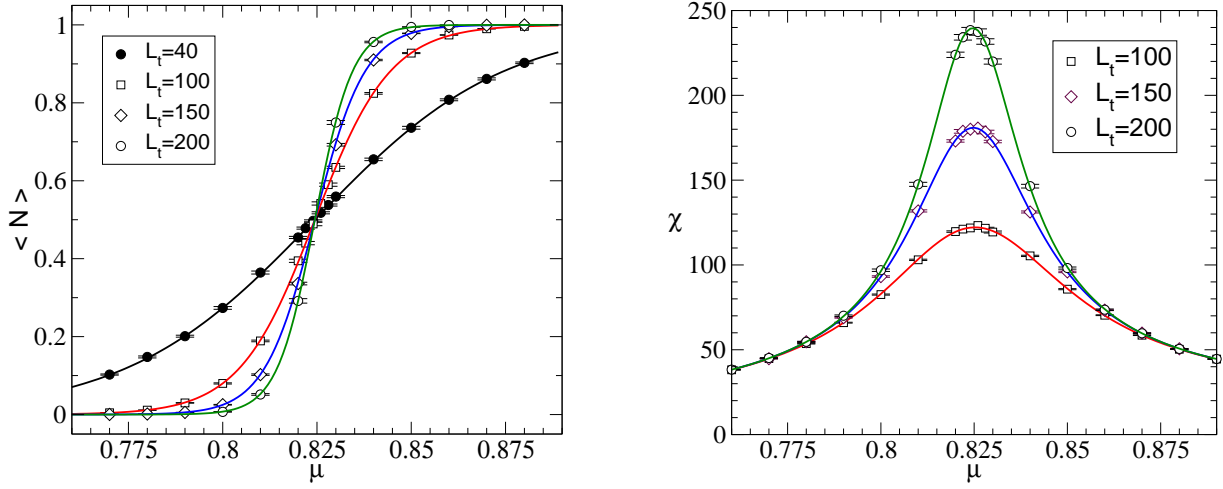

\begin{center}
\hbox{
\includegraphics[width=0.45\textwidth]{fig3a.eps}
\hskip0.5in
\includegraphics[width=0.45\textwidth]{fig3b.eps}
}
\end{center}
\vskip-0.5in
\caption{\label{fig3} The average particle number and the condensate susceptibility as a function of $\mu$ near the transition between $N=0$ and $N=1$ at $L=2$ and $\beta=0.43$ for different values of $L_t$. The solid lines show the fit of data to the effective quantum mechanics description discussed in the text.}
\end{figure*}

We first consider $L=2$ and vary $L_t$ in the range $40 \leq L_t \leq 200$ which is easy due to the small lattice size. In Fig.~\ref{fig2} we plot the behavior of our four observables as a function of the chemical potential at $L_t=100$ and $\beta=0.43$. Note that the particle number increases in steps of one at critical values of $\mu$. This means energy per particle of the ground state in every particle number sector increases with the number of particles. In other words the particles repel each other. Thus, small systems containing particles of the non-linear sigma model will show the phenomena similar to {\em Coulomb Blockade} observed in nanoscale systems \cite{QuantumDots}. By fitting the data at $L=2$ and $\beta=0.43,0.50$ and $0.20$ we have extracted the parameters $\mu_c^{(N)}$, $\rho_0$, $\rho_1$ and $\chi_0$ for $N=0,1,2$ and $3$. These are tabulated in Tab.~\ref{tab1}.  In order to show the goodness of our fits, in Fig.~\ref{fig3} we plot the behavior of $\langle N\rangle$ and $\chi$ for values of $\mu$ close to the transition between the $N=0$ and $N=1$ sector for different values of $L_t$ at $\beta=0.43$. The solid lines represent the fit functions using the parameter values given in table \ref{tab1}. Note that all the computed values of $\langle N \rangle$ shown in the left plot of Fig.~\ref{fig3} can be fit with just one parameter namely $\mu_c^{(N)}$. 

\begin{figure*}[b]
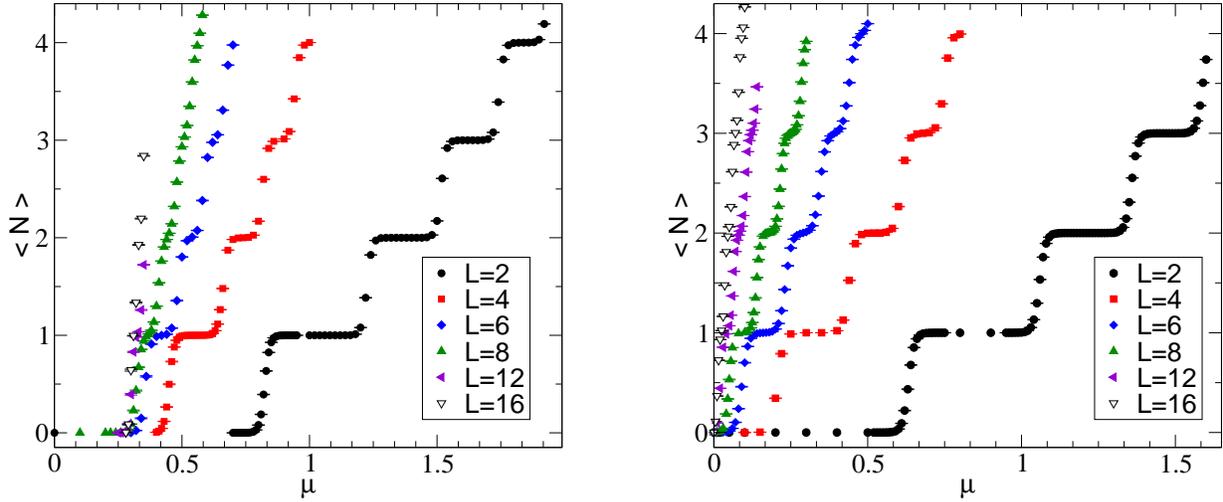

\begin{center}
\hbox{
\includegraphics[width=0.45\textwidth]{fig4a.eps}
\hskip0.5in
\includegraphics[width=0.45\textwidth]{fig4b.eps}
}
\end{center}
\vskip-0.5in
\caption{\label{fig4} The average particle number as a function of $\mu$ for at $\beta=0.43$ (left) $\beta=0.50$ (right) for different values of $L$. When $\beta=0.43$ the data shown is for $L_t=100$ at $L=2--8$, $L_t=200$ at $L=12$ and $L_t=300$ at $L=16$. When $\beta=0.50$ the data shown is for $L_t=100$ at $L=2--6$, $L_t=160$ at $L=8$, $L_t=200$ at $L=12$ and $L_t=300$ at $L=16$.}
\end{figure*}

\begin{figure*}
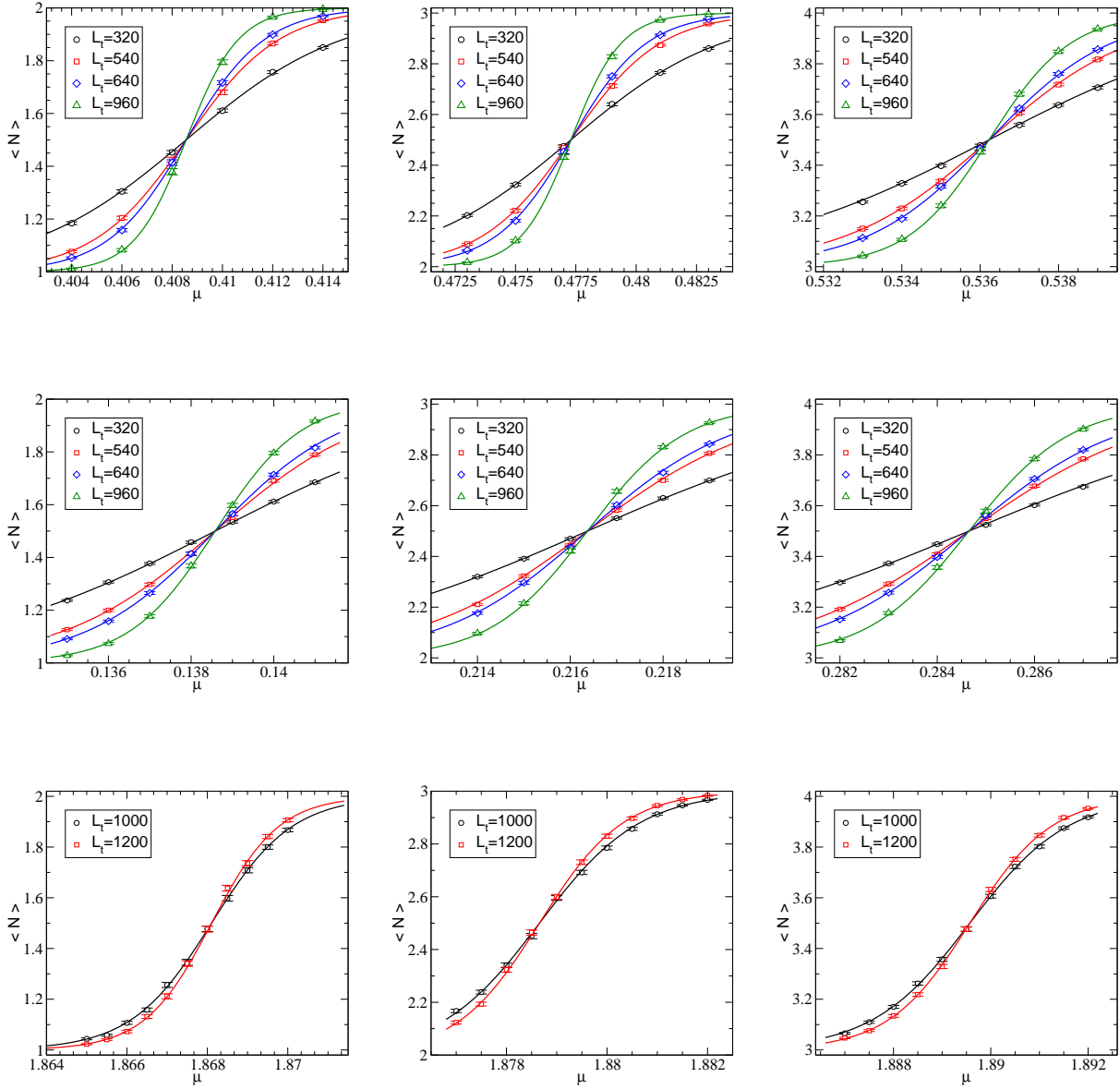

\begin{center}
\vbox{
\hbox{
\includegraphics[width=0.3\textwidth]{fig5a.eps}
\hskip0.2in
\includegraphics[width=0.3\textwidth]{fig5b.eps}
\hskip0.2in
\includegraphics[width=0.3\textwidth]{fig5c.eps}
}
\vskip0.5in
\hbox{
\includegraphics[width=0.3\textwidth]{fig5d.eps}
\hskip0.2in
\includegraphics[width=0.3\textwidth]{fig5e.eps}
\hskip0.2in
\includegraphics[width=0.3\textwidth]{fig5f.eps}
}
\vskip0.5in
\hbox{
\includegraphics[width=0.3\textwidth]{fig5g.eps}
\hskip0.2in
\includegraphics[width=0.3\textwidth]{fig5h.eps}
\hskip0.2in
\includegraphics[width=0.3\textwidth]{fig5i.eps}
}
}
\end{center}
\caption{\label{fig5} The average particle number as a function of $\mu$ near the transitions between $N$ and $N+1$ for $N=1,2,3$ at $L=8$ and $\beta=0.43$ (top), $\beta=0.5$ (center) and $\beta=0.2$ (bottom).The solid lines are fits to the data.}
\end{figure*}

\begin{figure*}
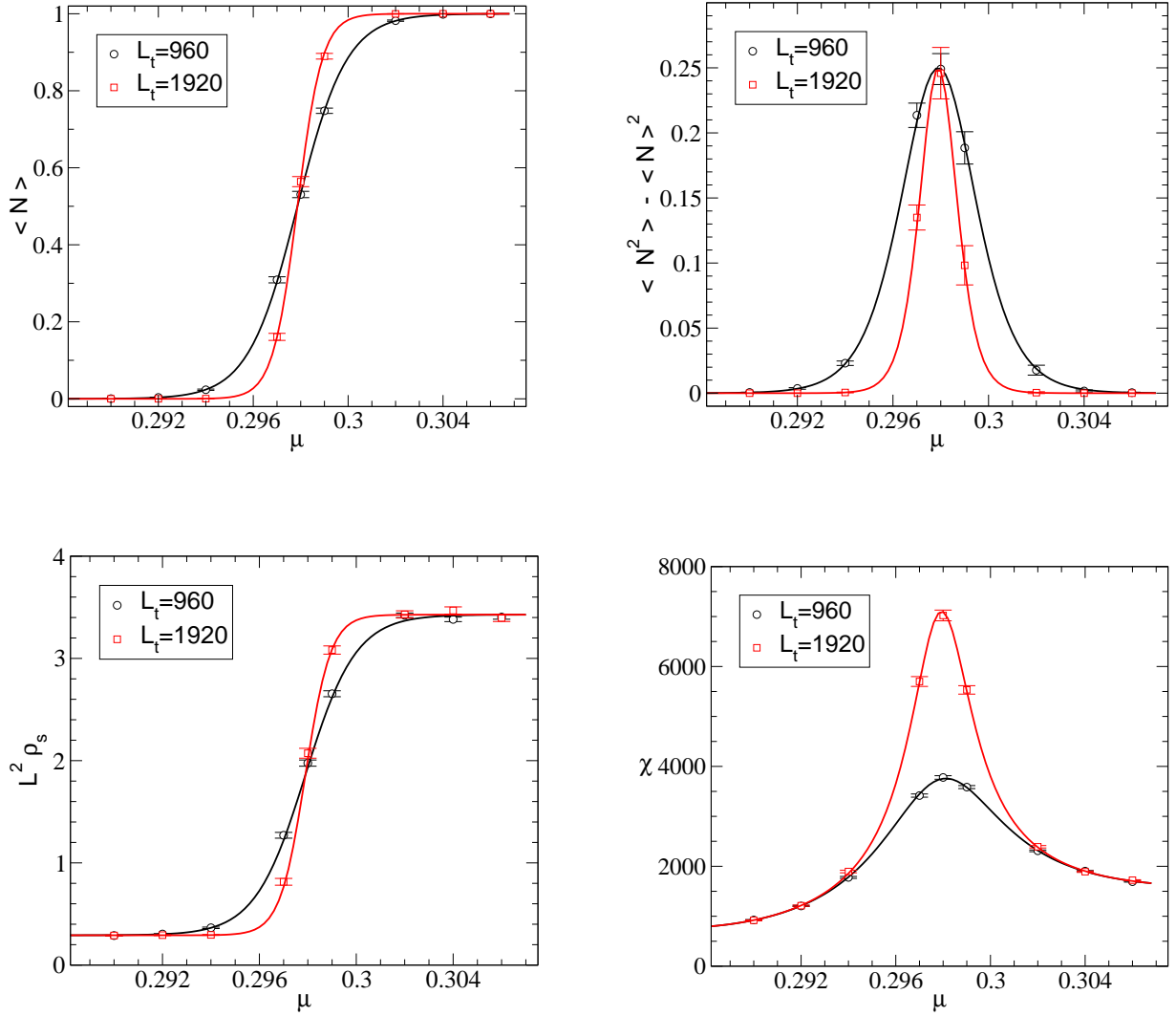

\begin{center}
\hbox{
\includegraphics[width=0.44\textwidth]{fig6a.eps}
\hskip0.5in
\includegraphics[width=0.46\textwidth]{fig6b.eps}
}
\vskip0.5in
\hbox{
\includegraphics[width=0.45\textwidth]{fig6c.eps}
\hskip0.5in
\includegraphics[width=0.45\textwidth]{fig6d.eps}
}
\end{center}
\caption{\label{fig6} The four observables as a function of $\mu$ near the transition between $N=0$ and $N=1$ at $L=16$ and $\beta=0.43$ at two different values of $L_t$. The solid lines show the fit of data to the effective quantum mechanics description discussed in the text.}
\end{figure*}

\begin{figure*}[b]
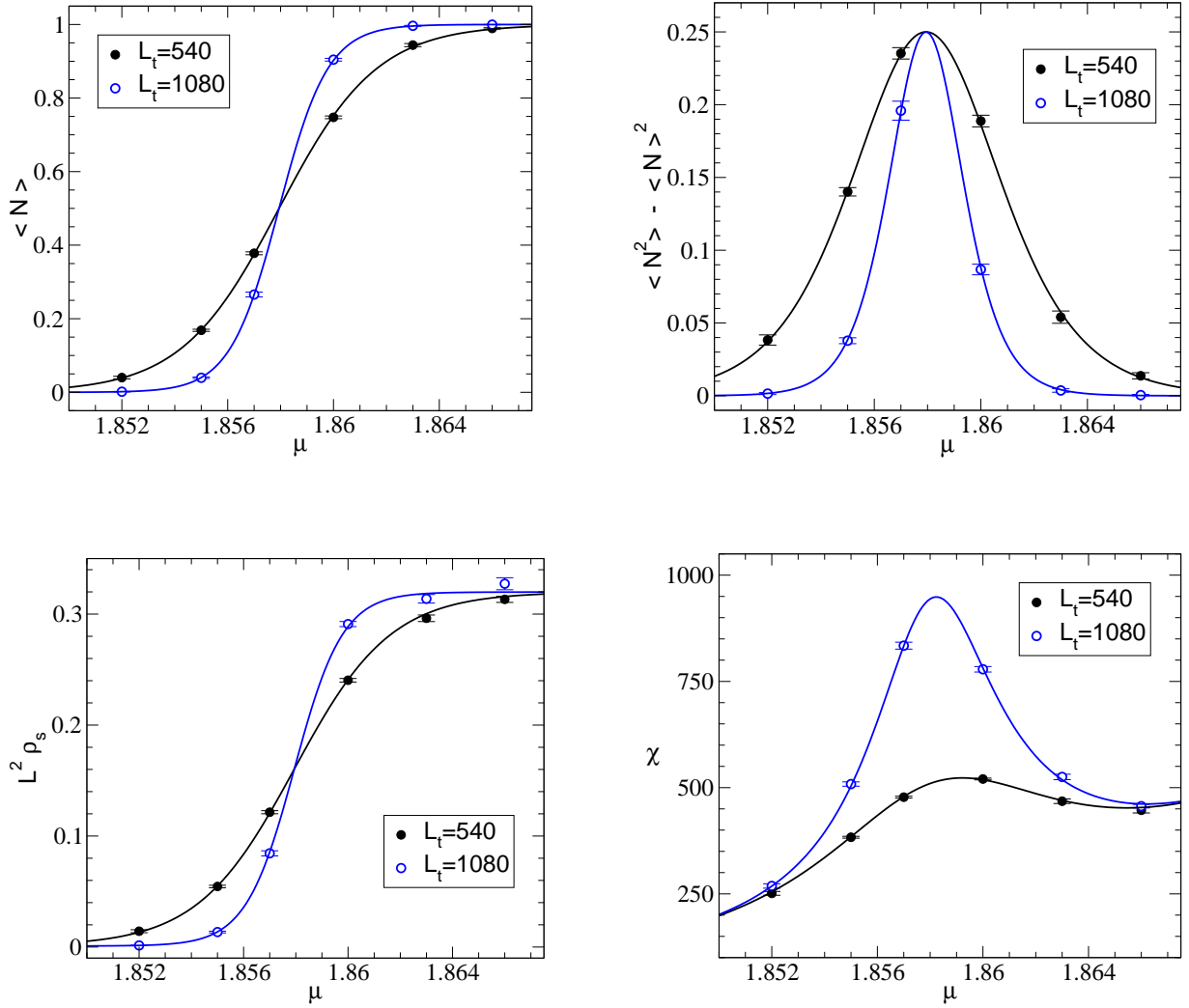

\begin{center}
\hbox{
\includegraphics[width=0.44\textwidth]{fig7a.eps}
\hskip0.5in
\includegraphics[width=0.46\textwidth]{fig7b.eps}
}
\vskip0.5in
\hbox{
\includegraphics[width=0.45\textwidth]{fig7c.eps}
\hskip0.5in
\includegraphics[width=0.45\textwidth]{fig7d.eps}
}
\end{center}
\caption{\label{fig7} The four observables as a function of $\mu$ near the transition between $N=0$ and $N=1$ at $L=6$ and $\beta=0.2$ for $L_t=540$ and $1080$. The solid lines show the fit of data to the effective quantum mechanics description discussed in the text.}
\end{figure*}

We have repeated the above analysis at larger values of $L$. We find the physics remains qualitatively similar to the $L=2$ case. In particular the average particle number jumps by one at critical values of $\mu$. In Fig.~\ref{fig4} we show the average particle number as a function of $\mu$ at different values of $L$ at $\beta=0.43$ and $\beta=0.50$. In Fig.\ref{fig5} we plot the average particle number at $L=8$ for all three values of $\beta$ at different values of $L_t$ close to $\mu_c^{(N)}$ for $N=1,2,3$. The effective quantum mechanics description continues to fit all our data well as long as $\mu$ is close to the critical values and $L_t$ is sufficiently large. The corresponding effective parameters are tabulated in Tabs.~\ref{tab2}, \ref{tab3} and \ref{tab4}. The fits always give reasonable $\chi^2/DOF$, which are shown in the last four columns, one for each observable. We note that as $\beta$ becomes smaller, $\mu_c^{(0)}$ becomes larger while $\mu_c^{(1)}-\mu_c^{(0)}$ becomes smaller. This is the reason it becomes difficult to match the data to an effective quantum mechanics description at small $\beta$ without going to very large $L_t$. Note also that the value of $\mu_c^{(0)}$ has approximately reached the thermodynamic limit at $\beta=0.2$ for $L=8$. We plot our four observables near $\mu_c^{(0)}$ at $L=16$ and $\beta=0.43$ in Fig.~\ref{fig6} and at $L=6$ and $\beta=0.2$ in Fig.~\ref{fig7} along with the fits.

\begin{table}[h]
\begin{tabular}{|c|c|c|c|c|c|c|c|c|c|c|c|c|c|}
\hline
\hline
$N$ & $L$ & $\mu_c^{N}$ & $\rho_0$ & $\rho_1$ & $\chi_0$ & \multicolumn{4}{|c|}{$\chi^2/DOF$} \\
\hline
0 & 4 & 1.85850(6)  & 0.00100(3)  & 0.02062(7) & 1.55(2)  & 1.40 & 1.50  & 1.42 & 0.85 \\
0 & 6 & 1.85794(2)  & 0.00002(1)  & 0.00888(3) & 1.62(2)  & 0.72 & 1.05  & 0.97 & 0.98 \\          
0 & 8 & 1.85798(3)  & 0.000001(1) & 0.00500(4) & 1.5(2)  & 0.58 & 0.56  & 0.92 & 0.39 \\
0 & 12 & 1.85801(3) & - & - & -  & 1.30 & -  & - &   \\ 
\hline
\hline
1 & 4 & 1.90898(5) & 0.02018(8)& 0.03768(9)& 2.98(3) &  1.44 & 1.61 & 1.64 & 1.41 \\
1 & 6 & 1.87771(2) & 0.00881(4)& 0.01731(6)& 3.04(6) &  1.07 & 0.83 & 0.56 & 0.81 \\
1 & 8 & 1.86810(1) & 0.00497(3)& 0.00987(5)& 3.1(1)  &  0.82 & 0.77 & 0.74 & 1.32 \\
\hline
\hline
2 & 4 & 1.95988(4) & 0.0371(1) & 0.0515(1) & 4.05(3) &  1.51 & 1.31 & 1.62 & 1.44\\
2 & 6 & 1.89811(2) & 0.01732(6)& 0.02494(7)& 4.33(6) &  0.91 & 0.82 & 1.53 & 0.48\\
2 & 8 & 1.87866(1) & 0.00986(6)& 0.01446(4)& 4.4(2)  &  0.61 & 0.59 & 0.63 & 0.43\\
\hline
\hline
3 & 4 & 2.01050(4) & 0.0512(1) & 0.0631(1) & 5.00(4) &  1.53 & 1.46 & 0.97 & 1.40\\
3 & 6 & 1.91911(1) & 0.02486(7)& 0.0323(1) & 5.65(7) &  1.59 & 1.38 & 0.92 & 0.40\\
3 & 8 & 1.88957(1) & 0.01451(5)& 0.01897(6)& 5.7(2)  &  0.58 & 0.55 & 0.74 & 0.42 \\
\hline
\end{tabular}
\caption{\label{tab2} Effective Quantum Mechanics parameters near the various particle number transitions from $N$ to $N+1$ at $\beta=0.20$ and various values of the spatial size $L$.}
\end{table}

\clearpage

\begin{table}[ht]
\begin{tabular}{|c|c|c|c|c|c|c|c|c|c|c|c|c|}
\hline
\hline
$N$ & $L$ & $\mu_c^{N}$ & $\rho_0$ & $\rho_1$ & $\chi_0$ & \multicolumn{4}{|c|}{$\chi^2/DOF$}\\
\hline
\hline
0 & 4 &  0.45019(3) &  0.0733(6) & 0.1307(5)& 4.50(3) &  0.42 &  0.50 &  0.92 &  1.00 \\
0 & 6 &  0.35692(3) &  0.0330(2) & 0.0766(3)& 5.83(5) &  0.34 &  0.31 &  0.55 &  1.21 \\
0 & 8 &  0.32270(3) &  0.0163(1) & 0.0489(2)& 6.3(1) &  0.94 &  1.01 &  0.60 &  0.42 \\
0 & 10&  0.30837(2) &  0.0081(2) & 0.0334(3)& 6.5(2) &  0.39 &  0.43 &  0.64 &  0.99 \\
0 & 12&  0.30207(2) &  0.0043(3) & 0.0236(4)& 7.1(2)  &  0.54 &  0.52 &  0.84 &  0.84 \\
0 & 16&  0.29789(2) &  0.0013(4) & 0.0135(4)& 7.0(4)  &  0.52 &  0.52 &  0.50 &  0.48 \\    
0 &  20&  0.29704(7) &  0.00039(5)& 0.0083(2)& 7.5(6) &  0.90 &  0.80 &  1.15 &  0.36 \\ 
\hline
\hline
1 & 4 & 0.66047(2) &  0.1311(6) & 0.1670(6) & 6.07(3) & 1.04 &  1.10 &  0.66 & 0.73 \\  
1 & 6 & 0.48596(2) &  0.0763(3) & 0.1060(3) & 8.56(6) & 0.63 &  0.63 &  1.23 & 0.57 \\
1 & 8 & 0.40856(2) &  0.0491(2) & 0.0728(2) & 10.2(1)&  0.38 &  0.36 &  1.13 & 0.43 \\
1 & 10& 0.36818(2) &  0.0336(2) & 0.0525(4) & 11.3(2) &  0.71 &  0.84 &  0.80 & 1.55 \\
1 & 12& 0.34516(1) &  0.0242(2) & 0.0387(2) & 11.9(2) &  0.59 &  0.57 &  0.90 & 0.51 \\
1 & 16& 0.32214(1) &  0.0136(2) & 0.0236(2) & 12.1(4) &  0.59 &  0.60 &  0.80 & 0.90 \\
\hline
\hline
2 & 4 &  0.81586(2) &  0.1676(6)&  0.1944(6)& 7.10(3) &  0.54 &  0.52 &  0.94 &  1.00 \\
2 & 6 &  0.58498(1) &  0.1057(3)&  0.1295(3)& 10.35(6)&  0.86 &  0.82 &  0.78 &  0.74 \\
2 & 8 &  0.47730(1) &  0.0727(2)&  0.0922(2)& 12.9(1)&  0.65 &  0.68 &  0.65 &  0.74 \\
2 & 10&  0.41819(1) &  0.0522(3)&  0.0686(3)& 14.3(2) &  0.70 &  0.66 &  1.18 &  0.76 \\
2 & 12&  0.38263(1) &  0.0391(3)&  0.0525(3)& 16.1(3) &  0.78 &  0.75 &  1.06 &  0.89 \\
2 & 16&  0.34468(1) &  0.0233(2)&  0.0335(2)& 17.7(5) &  0.39 &  0.37 &  0.78 &  0.75 \\     
\hline
\hline
3 & 4 & 0.94318(2) &  0.1944(7)& 0.2150(5)& 7.94(3) &  0.50 &  0.53 &  0.92 &  1.10 \\
3 & 6 & 0.66810(1) &  0.1293(3)& 0.1478(3)& 11.93(6)&  0.54 &  0.87 &  0.99 &  0.48 \\
3 & 8 & 0.536227(9)&  0.0916(2)& 0.1079(3)& 15.1(1)&  0.47 &  0.47 &  1.24 &  0.98 \\  
3 & 10& 0.46209(1) &  0.0679(3)& 0.0820(3)& 17.6(2) &  0.58 &  0.57 &  0.93 &  0.59 \\    
3 & 12& 0.41633(1) &  0.0526(3)& 0.0637(4)& 19.2(3) &  0.46 &  0.44 &  0.80 &  0.37 \\
3 & 16& 0.365744(9)&  0.0328(3)& 0.0412(2)& 22.8(5) &  0.80 &  0.76 &  0.58 &  1.08 \\
\hline
\hline
\end{tabular}
\caption{\label{tab3} Effective Quantum Mechanics parameters near the various particle number transitions from $N$ to $N+1$ at $\beta=0.43$ and various values of the spatial size $L$.}
\end{table}

\begin{table}[t]
\begin{tabular}{|c|c|c|c|c|c|c|c|c|c|c|c|c|}
\hline
\hline
N & $L$ & $\mu_c^{N}$ & $\rho_0$ & $\rho_1$ & $\chi_0$ & \multicolumn{4}{|c|}{$\chi^2/DOF$} \\
\hline
\hline
0 & 4 & 0.20651(2) & 0.1868(7)& 0.2173(7)& 7.01(3) & 0.62 & 0.59 & 0.72 & 1.18\\
0 & 6 & 0.09166(2) & 0.1717(5)& 0.1847(5)& 13.39(9)&  0.69 & 0.65 & 1.21 & 0.36\\
0 & 8 & 0.049211(9)& 0.1691(5)& 0.1732(5)& 22.0(1) &  0.82 & 0.79 & 1.08 & 0.80\\
0 & 10& 0.030691(8)& 0.1677(7)& 0.1700(6)& 32.5(3) &  0.54 & 0.51 & 0.40 & 0.64\\
0 & 12& 0.02112(1) & 0.1677(4)& 0.1681(4)& 45.1(7) &  0.63 & 0.60 & 1.06 & 1.10\\
0 & 16& 0.011786(8)& 0.1670(4)& 0.1670(4)& 77.3(8) &  0.68 & 0.63 & 0.88 & 0.74\\
\hline
\hline
1 & 4 & 0.43892(2) & 0.2162(7) & 0.2440(7) & 7.76(3)  &  0.87 & 0.83 & 0.30 & 0.68\\
1 & 6 & 0.23271(2) & 0.1832(5) & 0.1997(5) & 14.00(9) &  0.42 & 0.40 & 0.47 & 1.12\\
1 & 8 & 0.138584(9)& 0.1737(4) & 0.1801(5) & 22.3(1)  &  0.72 & 0.69 & 1.23 & 0.82\\
1 & 10& 0.090004(8)& 0.1699(6) & 0.1729(6) & 32.9(3)  &  0.77 & 0.74 & 0.63 & 1.5\\
1 & 12& 0.06268(1) & 0.1680(4) & 0.1704(5) & 45.6(7)  &  1.50 & 1.42 & 0.52 & 0.66\\
1 & 16& 0.035256(8)& 0.1672(4) & 0.1673(4) & 77.0(8)  &  0.82 & 0.78 & 0.94 & 0.70\\
\hline
\hline
2 & 4 & 0.61012(2) & 0.2447(7) & 0.2638(7) & 8.40(3)  &  0.86 & 0.83 & 1.47 & 1.05\\
2 & 6 & 0.34520(1) & 0.1986(9) & 0.2131(9) & 14.99(8) &  0.32 & 0.30 & 1.11 & 0.84\\
2 & 8 & 0.216371(9)& 0.1813(6) & 0.1891(6) & 23.1(1)  &  0.79 & 0.75 & 0.79 & 0.82\\
2 & 10& 0.145056(7)& 0.1729(6) & 0.1789(6) & 33.4(3)  &  0.88 & 0.83 & 1.12 & 0.94\\
2 & 12& 0.10266(1) & 0.1701(4) & 0.1734(4) & 44.8(7)  &  0.72 & 0.68 & 0.49 & 1.02\\
2 & 16& 0.058435(8)& 0.1680(4) & 0.1687(4) & 78.4(8)  &  0.34 & 0.32 & 0.82 & 0.85\\
\hline
\hline
3 & 4 & 0.74886(1) & 0.2639(7) & 0.2824(7) & 8.98(3)  &  0.72 & 0.69 & 0.90 & 0.46\\
3 & 6 & 0.43967(1) & 0.2123(9) & 0.2280(9) & 15.74(8) &  0.43 & 0.40 & 1.16 & 0.89\\
3 & 8 & 0.284653(8)& 0.1900(6) & 0.1975(6) & 24.1(1)  &  0.72 & 0.69 & 1.25 & 0.79\\
3 & 10& 0.195544(7)& 0.1780(7) & 0.1838(6) & 34.8(3)  &  0.63 & 0.60 & 1.09 & 0.93\\
3 & 12& 0.14064(1) & 0.1729(4) & 0.1762(4) & 46.1(7)  &  1.04 & 0.98 & 0.58 & 1.50\\
3 & 16& 0.081185(8)& 0.1689(4) & 0.1698(4) & 78.3(8)  &  0.88 & 0.83 & 0.66 & 1.39\\
\hline
\end{tabular}
\caption{\label{tab4} Effective Quantum Mechanics parameters near the various particle number transitions from $N$ to $N+1$ at $\beta=0.50$ and various values of the spatial size $L$.}
\end{table}

\clearpage

\clearpage

\section{Thermodynamic Limit}
\label{res2}

Using the results of the previous section it is tempting to extrapolate to the thermodynamic limit. However, in order to accomplish this task it is important to know how the effective quantum mechanical parameters depend on $L$. This dependence is non-universal in general and close to a critical point will depend on the nature of the phase transition. Assuming the phase transition is second order, close to the critical chemical potential where the density can be made arbitrarily small, we expect universal features to emerge. For example, when the particles have a purely repulsive interaction, the ground state energy of $N$ particles is always less than the corresponding energy of $N+1$ particles \cite{PhysRev.116.1344}. Based on the results of the previous section this scenario seems to be valid in the current model. Indeed the particle number always increases by one as we increase $\mu$ at every fixed value of $L$. The superfluid density $\rho_s$ also behaves like $\rho$. Thus, we conclude that at $\mu=\mu_c^{(0)}$ in the thermodynamic limit, there is a second order transition to a superfluid phase. Based on this, below we discuss the extrapolations to the thermodynamic limit.

\begin{figure*}[t]
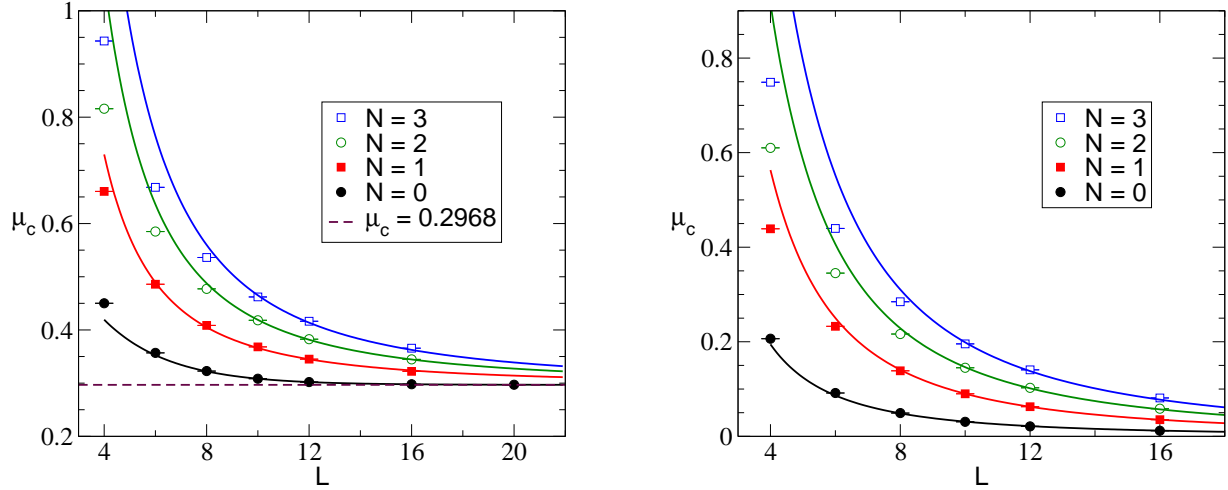

\begin{center}
\vbox{
\includegraphics[width=0.45\textwidth]{fig8a.eps}
\hskip0.5in
\includegraphics[width=0.45\textwidth]{fig8b.eps}
}
\end{center}
\caption{\label{fig8} The finite size scaling of the N-particle energy levels as a function of the spatial lattice size $L$ at $\beta=0.43$ (left, massive phase) and $\beta=0.50$ (right,superfluid phase).}
\end{figure*}

First we consider $\beta=0.43$ where the low energy physics contains massive bosons with repulsive interactions. Then, the quantity $\mu_c^{(0)}$ is simply the mass of the particle $M(L)$ at a finite $L$. This mass can be obtained by fitting the the temporal two-point correlation function
\begin{equation}
G(t) = \sum_{x_\perp,y_\perp} 
\Big\langle \mathrm{e}^{i\theta_{x_\perp}}\mathrm{e}^{-i\theta_{y_\perp}}\Big\rangle,
\end{equation}
computed at $\mu=0$, to the form $G(t) \sim \exp(-M(L) t)$ for values of $t \ll L_t/2$. In the definition of $G(t)$, $y_\perp$ and $x_\perp$ represent lattice sites at temporal slices $0$ and $t$ respectively. We have computed $M(L)$ using this method and indeed we find excellent agreement with $\mu_c^{(0)}$ at all the three values of $\beta$. This means the true mass of the particle must be
\begin{equation}
M = \lim_{L \rightarrow \infty} \mu_c^{(0)}.
\end{equation}
We can reverse this argument and obtain $\mu_c^{(0)}$ in the thermodynamic limit by simply measuring the mass of the particle at $\mu=0$. Of course this result is not general and is valid only in the present study where there is clear evidence that the particles repel each other. In order to extract $M$ in the massive phase ($\beta < \beta_c$) we can use L\"{u}scher's formula \cite{Luscher:1985dn} extended to two spatial dimensions,
\begin{equation}
\mu_c^{0} \approx M + M_1\mathrm{e}^{-\tilde{m} L}.
\label{muc1}
\end{equation}
At $\beta=0.43$ we find that $\mu_c^{(0)}$ fits well to this form and gives $M=0.29680(4)$, $M_1=0.59(2)$ and $\tilde{m} = 0.393(4)$ with a small $\chi^2/DOF$. The data and the fit are shown in the left plot of Fig.~\ref{fig8}.

The spatial size dependence of the energy of $N$ particles in three spatial dimensions has been calculated using models of quantum mechanics \cite{Bogolyubov:1947zz,Huang:1957im}. Recently, this dependence was also computed using effective field theory \cite{Beane:2007qr}. In the special case of two particles the problem was also solved in a general massive quantum field theory in three spatial dimensions \cite{Luscher:1986pf}. All these studies indicate that the ground state energy of $N$ particles satisfies the relation $E_0^{(N)}-E_0^{(0)} \propto N (N+1)/L^3$.  For $N \geq 1$, remember that $\mu_c^{(N)}$ is the difference in the ground state energies of $N+1$ particles and $N$ particles. Extending the known results to two spatial dimensions and making the assumption that the particle density in the thermodynamic limit close to the critical point is of the form $\rho \sim c (\mu - \mu_c^{(0)})$ at leading order in the superfluid phase, we expect
\begin{equation}
\mu_c^{(N)}(L) = \mu_c^{(0)} + \frac{(N+1)}{c L^2}
\label{mucN}
\end{equation}
for sufficiently large $L$ and $N$. Figure \ref{fig8} shows that our data is described reasonably well by this equation. In the left plot of Fig.~\ref{fig8} we show the values of $\mu_c^{(N)}(L)$ obtained from table \ref{tab3}. The solid lines show the dependence of $\mu_c^{(N)}(L)$ on $L$ as described by eq.~(\ref{mucN}) with $c \approx 0.18, 0.16$ and $0.15$ at $N=1,2$ and $3$ respectively. Clearly, for large values of $L$ the solid lines pass through the data. The value of $c$ changes slightly since $N$ is small. Unfortunately a fit of our data to eq.~(\ref{mucN}) yields a large $\chi^2/DOF$. We believe this is due to the fact that our data has very small errors and hence is sensitive to higher order corrections which we do not know analytically at the moment in two spatial dimensions.

When $\beta = 0.5$ we are in the superfluid phase and the $U(1)$ particle number symmetry is spontaneously broken. One then expects the low energy spectrum at finite volumes to be governed by $O(2)$ chiral perturbation theory. Based on this we again expect $\mu_c^{(N)}(L)$ to be described by eq.~(\ref{mucN}) but with $\mu_c^{(0)}=0$. While our data is again consistent with these expectations (see right plot of Fig.~\ref{fig8}), without keeping higher order $1/L$ corrections the fits again give a large $\chi^2/DOF$. The solid lines in Fig.~\ref{fig8} describe eq.~(\ref{mucN}) with $c = 0.33, 0.22, 0.2, 0.2$ for $N=0,1,2,3$.

\section{Phase Diagram}
\label{pdiag}

\begin{figure}[t]
\begin{center}
\includegraphics[width=0.58\textwidth]{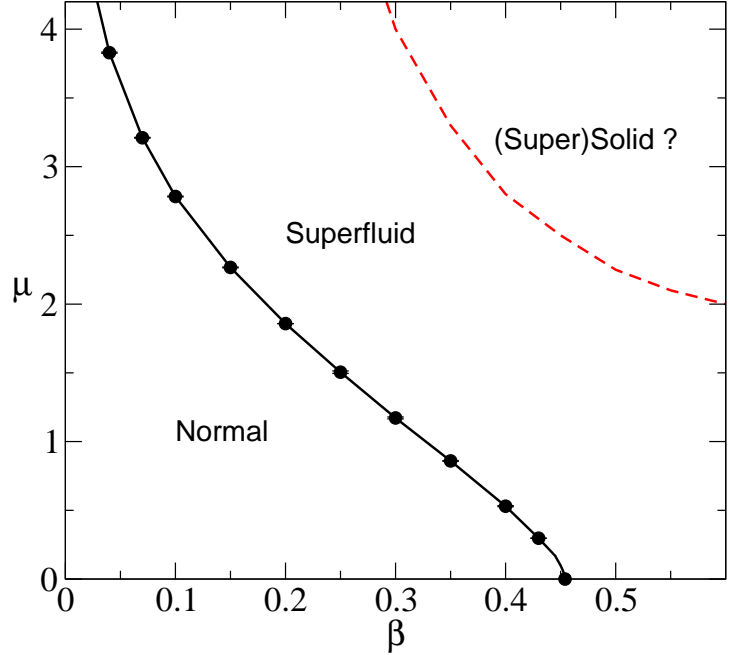}
\end{center}
\caption{\label{fig9} The phase diagram in the $\beta$ vs. $\mu$ plane. The circles show the value of $\mu_c^{(0)}$ as a function of $\beta$ given in table \ref{tab5}. The solid line that connects these points forms the phase boundary between the normal phase and the superfluid phase. This transition is second order. Since the particles repel each other we speculate that at higher densities a first order transition (dashed line) may separate the superfluid phase from a solid or a super-solid phase.}
\end{figure}

The phase diagram of the $O(2)$ non-linear sigma model is an interesting research topic in itself. While the complete phase diagram requires more work, our results above allow us to compute the location of the transition line between the normal phase and the superfluid phase. In particular the value of $\mu_c^{(0)}$ as a function of $\beta$ determines this line. Based on the evidence at $\beta=0.43$ and $0.20$ we predict that $\mu_c^{(0)} = M$ for all values of $\beta < \beta_c$. The coordinates of the transition line are tabulated in Tab.~\ref{tab5} and shown on the phase diagram in Fig.~\ref{fig9}. We expect this transition to be second order in the mean field universality class with logarithmic corrections except at $\mu=0$ where it is governed by the $3d$ XY universality class. Thus, when $\beta <\beta_c$ and $(\beta_c - \beta)/\beta_c \ll 1$ we must have $\mu_c^{0} \propto [(\beta_c-\beta)/\beta_c]^\nu$ where $\nu \approx 0.671$ \cite{Campostrini:2000iw}.

\begin{table}[h]
\begin{tabular}{c|c|c|c|c|c|c|c|c|c|c}
\hline
$\beta$ & 0.43 & 0.40 & 0.35 & 0.30 & 0.25 & 0.2 & 0.15 & 0.10 & 0.07 & 0.04\\
\hline
$\mu_c^{(0)}$ & 0.29678(3) & 0.530(4) & 0.859(2) & 1.172(7) & 1.505(9) & 1.85801(3) & 2.267(3) & 2.783(3) & 3.210(3) & 3.829(4) \\
\hline
\end{tabular}
\caption{\label{tab5} The values of $\mu_c^{(0)}$ obtained by assuming that it is equal to the $L\rightarrow \infty$ limit of $M(L)$ at $\mu=0$ as discussed in the text.}
\end{table}

In principle there could be other interesting phases at larger values of $\mu$ which we cannot rule out based on the current work. Since we have seen the particles have a repulsive interaction, an interesting possibility is the existence of a solid phase or a super-solid phase \cite{supersolid}. However, there are stringent constraints for super-solids to arise \cite{PhysRevLett.25.1543,PhysRevLett.94.155302} and we do not know if these rule out such a phase in the current model. In any case if a transition to a solid phase exists, it will most likely be first order similar to the solid-liquid phase transitions in materials and will occur at densities where the lattice structure may become important. These transitions can also be studied efficiently with the worm algorithm. We postpone these studies for the future, but have speculated the possibility of a solid phase in Fig.~\ref{fig9}.

\section*{Acknowledgments}

We would like to thank Jyotirmoy Maiti and Sayantan Sharma for contributing to this project in its initial stages. DB would like to thank Arnab Sen for discussions at various stages of the project. SC would like to thank the members of the theory group at the Tata Institute of Fundamental Research (TIFR) for invitation and hospitality where this work was envisioned and partly accomplished. We would like to thank Gert Aarts for motivating us in this work and Philippe de Forcrand for a useful observation. This work was supported in part by the U.S. Department of Energy (USDOE) grant DE-FG02-05ER41368 and the U.S. National Science Foundation grant DMR-0506953. The computational resources used were two computer clusters, one funded by the USDOE located at Duke University and the other funded by the DAE of India located at TIFR.

\clearpage

\bibliography{algo}

\newpage

\appendix

\section{The Worm Algorithm}
\label{algo}

The worm algorithm for the partition function described by eq.~\ref{wlpart} can easily be constructed using ideas from \cite{Prokof'ev:2001zz,Chandrasekharan:2006tz,Wolff:2009kp}. Here we outline the essential steps of the update for completeness. Each worm update is as follows
\begin{enumerate}
\item We pick a random point $x$ on the lattice. We will also call this site $x_{\rm first}$. We set a counter $c=0$.
\item We pick at random one of $2d$ neighbors $x+\hat{\alpha},\alpha = \pm 1,\pm 2,...,\pm d $ of the site $x$.
\item Let $k$ be the current on the bond connecting $x$ and $x+\alpha$. If $\alpha$ is positive then with probability
\[
\frac{I_{k+1}(\beta)\mathrm{e}^{\mu \delta_{\alpha,t}}}{I_k(\beta)}
\]
we change $k$ to $k+1$ and move to the neighboring site $x+\hat{\alpha}$. If $\alpha$ is negative then with probability
\[
\frac{I_{k-1}(\beta)\mathrm{e}^{-\mu \delta_{\alpha,t}}}{I_k(\beta)}
\]
we change $k$ to $k-1$ and move to the neighboring site $x+\hat{\alpha}$. Otherwise we stay at site $x$.
\item We set $c = c+1$. If $x=x_{\rm first}$ we stop and complete one worm update. Otherwise we go to step 2 and repeat the above steps.
\end{enumerate}
It turns out that $\chi$ is given by the average of $c$ after many worm updates. The other observables are measured on each world-line configuration and averaged over the ensemble generated by the worm algorithm.

\begin{table}
\begin{center}
 \begin{tabular}{|c|c|c|c|c|c|c|c|c|}
 \hline
$\beta$&$\chi$&$\chi^{MC}$&$\rho$&$\rho^{MC}$&$\kappa$&$\kappa^{MC}$&$\rho_s$&$\rho_s^{MC}$\\
\hline
\multicolumn{9}{|c|}{$\mu=0.0$} \\
\hline
$0.1$&$1.2207$&$1.2206(1)$&$0$&$2(2) \times 10^{-5}$&$0.01005$&$0.01001(4)$&$0.01005$&$0.01008(4)$\\
$0.2$&$1.4831$&$1.4830(2)$&$0$&$-5(4)\times10^{-5}$&$0.04064$&$0.04063(8)$&$0.04064$&$0.04076(8)$\\
$0.5$&$2.3838$&$2.3835(4)$&$0$&$1(9)\times10^{-5}$&$0.2526$&$0.2527(2)$&$0.2526$&$0.2526(2)$\\
$1.0$&$3.2730$&$3.2733(4)$&$0$&$2(2)\times10^{-4}$&$0.7796$&$0.7796(4)$&$0.7796$&$0.7794(4)$\\
$5.0$&$3.8728$&$3.8733(3)$&$0$&$1(1)\times10^{-3}$&$4.809$&$4.806(4)$&$4.809$&$4.813(4)$\\
\hline
\hline
\multicolumn{9}{|c|}{$\mu=0.5$} \\
\hline
$0.1$&$1.2362$&$1.2361(1)$&$0.00590$&$0.00592(2)$&$0.01559$&$0.01560(5)$&$0.01007$&$0.01013(4)$\\
$0.2$&$1.5188$&$1.5188(2)$&$0.02374$&$0.02378(5)$&$0.0640$&$0.0640(1)$&$0.04100$&$0.04105(7)$\\
$0.5$&$2.4692$&$2.4693(4)$&$0.1429$&$0.1430(1)$&$0.4341$&$0.4344(3)$&$0.2581$&$0.2581(2)$\\
$1.0$&$3.3245$&$3.3247(4)$&$0.4190$&$0.4193(2)$&$1.6553$&$1.6564(10)$&$0.7867$&$0.7861(4)$\\
\hline
\hline
\multicolumn{9}{|c|}{$\mu=1.0$} \\
\hline
$0.1$&$1.2861$&$1.2861(2)$&$0.01809$&$0.01812(4)$&$0.03851$&$0.03857(8)$&$0.01018$&$0.01024(4)$\\
$0.2$&$1.6316$&$1.6317(2)$&$0.07160$&$0.07162(8)$&$0.1637$&$0.1638(2)$&$0.04241$&$0.04242(8)$\\
$0.5$&$2.6899$&$2.6900(4)$&$0.3865$&$0.3868(2)$&$1.2530$&$1.2539(8)$&$0.2747$&$0.2748(2)$\\
$1.0$&$3.4317$&$3.4317(4)$&$1.0053$&$1.0055(3)$&$5.487$&$5.488(3)$&$0.8032$&$0.8033(4)$\\
\hline
\end{tabular}
\end{center}
\caption{Checks of the observables with exact solution on $2\times 2$ lattices.\label{chk1}}
\end{table}

\section{Tests of the Algorithm}
\label{algotest}

We have verified our algorithm and code by both solving the model exactly on a $2 \times 2$ lattice as well as comparing with the available results in the literature for $\mu=0$ in three dimensions. In this section we describe some of these tests. First, we compare the results of the various observables computed using the directed path algorithm with the exact results on a $2 \times 2$ lattice. The comparison is shown in Table~\ref{chk1}. Since space and time are symmetric we expect $\rho_s = \kappa$ at $\mu=0$. Our results reflect this fact.

\begin{figure}[t]
\begin{center}
\includegraphics[width=0.7\textwidth]{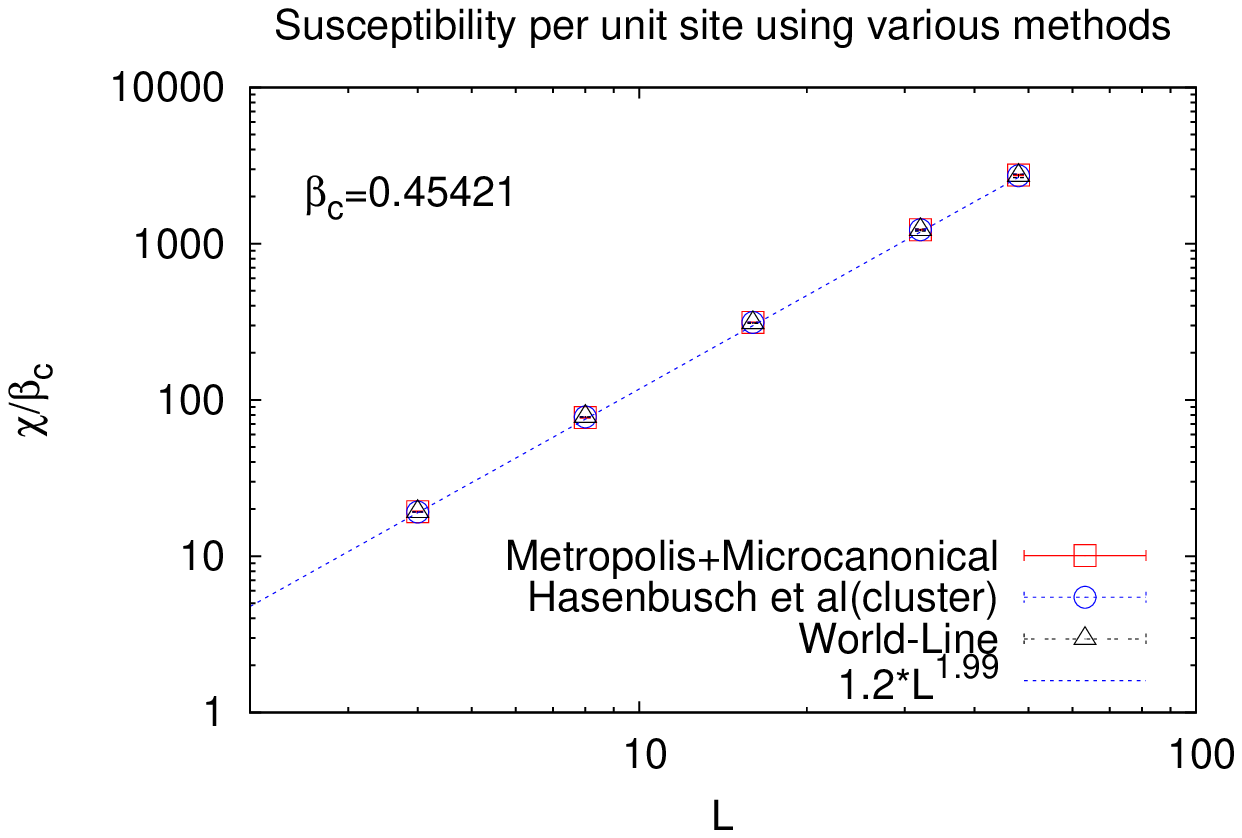}
\caption{Plot of $\chi$ vs $L$ at $\beta=\beta_c$. }
\label{fig:chk3}
\end{center}
\end{figure}

Extending the code from two dimensions to three dimensions is trivial and the chance for mistakes is rather small. However, we have tested the code at least at $\mu=0$ using the results from previous work. Here we compare results for $\chi$ obtained using the worm algorithm with that obtained using the microcanonical improved Metropolis (MM) algorithm and the available results in the literature using the cluster method \cite{Hasenbusch:1989fi}. The comparison is shown in Table~\ref{chk2}. The reason for us to choose $\beta=0.45421$ is because this is known to be the critical value of the coupling where the theory undergoes a phase transition from a normal phase to a superfluid phase. At the critical coupling we expect $\chi \propto L^{\gamma/\nu}$. A fit of our data to this form yields the value of $\gamma/\nu=1.99$ as expected from \cite{Hasenbusch:1989fi}, and is shown in figure \ref{fig:chk3}.

\begin{table}[b]
\begin{center}
\begin{tabular}{c|c|c|c|c}
\hline
$\beta$ & $L$ & Worm & MM & Cluster\\ 
\hline
$0.45421$ & $4$  &$19.17(3)$&$19.24(13)$  & $19.15(5)$\\
$0.45421$ & $8$  &$77.8(2)$&$76.9(5)$     & $77.9(3)$\\
$0.45421$ & $16$ &$310(1)$&$313(2)$       & $313(2)$\\
$0.45421$ & $32$ &$1221(18)$&$1228(7)$    & $1226(13)$\\
$0.45421$ & $48$ &$2713(67)$&$2750(27)$   & $2719(68)$\\
$0.01$    & $8$  & $1.0304(2)$&$1.03(46)$ & -\\
$0.1$     & $8$  & $1.3976(9)$&$1.40(6)$  & - \\
$1.0     $& $8$  & $387.2(3)$&$387.12(3)$ & - \\
\hline
\end{tabular}
\end{center}
\caption{Comparison of the condensate susceptibility $\chi$ with results from the worm algorithm, the Metropolis+Microcanonical(MM) update and Wolff Cluster update on $L^3$ lattices at different values of $L$ and $\beta$.\label{chk2} }
\end{table}

\clearpage

\section{Raw Data for comparison at $\mu\neq 0$}
\label{rawdata}

Here we give some raw data obtained using our algorithm at non-zero chemical potential for comparison with other methods like the complex Langevin method which are being developed to solve the sign problem present in the conventional formulation. These values can also serve as a check for future work.

\begin{table}[h]
\begin{center}
\begin{tabular}{|c|c|c|c|c|c|c|c|}
\hline
$\beta$ & $L$ & $L_t$ & $\mu$ & $\rho$ & $\kappa$ & $\rho_s$ & $\chi$ \\
\hline
0.5 & 12 & 144 & $0.035$ & $0.0625(1)$ & $0.932(2)$ & $0.1685(7)$ & $3960(11)$ \\
0.43 & 4 & 100 & $0.446$ & $0.0248(4)$ & $2.48(4)$ & $0.0972(8)$ & $239(2)$ \\
0.43 & 16 & 960 & $0.297$ & $0.00120(3)$ & $1.15(3)$ & $0.0050(1)$ & $3422(31)$ \\
0.30 & 8 & 8 & $1.20$ & $0.0450(3)$ & $1.39(1)$ & $0.0099(1)$ & $34.8(2)$ \\ 
0.30 & 8 & 64 & $1.20$ & $0.0246(2)$ & $3.17(3)$ & $0.0164(1)$ & $250(2)$ \\ 
0.20 & 6 & 180 & $1.855$ & $0.0104(2)$ & $1.94(4)$ & $0.00326(6)$ & $188(2)$ \\ 
\hline
\end{tabular}
\end{center}
\caption{Raw data obtained using the worm algorithm at random values of the parameters}
\end{table}

\begin{table}[h]
\label{}
\begin{center}
\begin{tabular}{c|c|c|c|c|c}
\hline
$L_t$ & $\mu$ & $\rho$ & $\kappa$ & $\rho_s$ & $\chi$ \\
\hline
   320 &    0.404 &     0.01850(8)  &        7.76(8)   &     0.0535(4)  &    1509(13)\\
   320  &   0.406  &    0.02038(9)    &      9.57(9)    &    0.0562(4)   &   1661(12)\\
   320   &  0.408  &    0.0227(1)    &        11.8(1)    &    0.0597(4)   &   1752(15)\\
   320   &   0.41   &    0.0252(1)     &       14.2(1)    &    0.0646(4)   &   1766(13)\\
   320  &   0.412  &    0.02745(8)    &      16.35(8)   &    0.0668(5)   &   1658(12)\\
   320  &   0.414   &   0.02890(6)    &      17.75(6)   &    0.0692(4)    &  1503(11) \\
\hline
\end{tabular}
\end{center}
\caption{Raw data at $\beta=0.43$, $L=8$ near the transition from N=1 to N=2.}
\end{table}

\end{document}